\documentclass{ws-rv975x65}
\usepackage{ws-rv-van}

\begin{document}

\chapter{Anderson localization and Supersymmetry\label{review}}
\author[K.B. Efetov]{K.B. Efetov}


\address{Theoretische Physik III, Ruhr-Universit\"at Bochum, 44780, Bochum,
Germany \\efetov@tp3.rub.de}

\begin{abstract}
The supersymmetry method for study of disordered systems is
shortly reviewed. The discussion starts with a historical
introduction followed by an explanation of the idea of using
Grassmann anticommuting variables for investigating disordered
metals. After that the nonlinear supermatrix $\sigma$-model is
derived. Solution of several problems obtained with the help of
the $\sigma$-model is presented. This includes the problem of the
level statistics in small metal grains, localization in wires and
films, and Anderson metal-insulator transition. Calculational
schemes developed for studying these problems form the basis of
subsequent applications of the supersymmetry approach.
\end{abstract}




\section{Introduction}
The prediction of the new phenomenon of the Anderson localization \cite%
{anderson} has strongly stimulated both theoretical and
experimental study of disordered materials. This work demonstrates
the extraordinary intuition of the author that allowed him to make
outstanding predictions. At the same time, one could see from that
work that quantitative description of the disordered systems was
not a simple task and many conclusions were based on
semi-qualitative arguments. Although many interesting effects have
been predicted in this way, development of theoretical methods for
quantitative study of quantum effects in disordered systems was
clearly very demanding.

The most straightforward way to take into account disorder is
using perturbation theory in the strength of the disorder
potential \cite{agd}. However, the phenomenon of the localization
is not easily seen within this method and the conventional
classical Drude formula for conductivity was considered in
\cite{agd} as the final result for the dimensionality $d>1$. This
result is obtained after summation of diagrams without
intersection of impurity lines. Diagrams with intersection of the
impurity lines give a small contribution if the disorder potential
is not strong, so that $ \varepsilon _{0}\tau \gg 1,$ where
$\varepsilon _{0}$ is the energy of the particles (Fermi energy in
metals) and $\tau $ in the elastic scattering time.

Although there was a clear understanding that the diagrams with
the intersection of the impurity lines were not small for one
dimensional chains, $d=1$, performing explicit calculations for
those systems was difficult. This step has been done considerably
later by Berezinsky \cite{berezinsky} who demonstrated
localization of all states in $1D$ chains by summing complicated
series of the perturbation theory. This result confirmed the
conclusion of Mott and Twose \cite{mott} about the localization in
such systems made previously. As concerns the higher dimensional
systems, $d>1$, the Anderson transition was expected at a strong
disorder but it was clear that the perturbation theory could not
be applied in that case.

So, the classical Drude theory was considered as a justified way
of the description of disordered metals in  $d>1$ and
$\varepsilon _{0}\tau \gg 1$.   At the same time, several results
for disordered systems could not be understood within this simple
generally accepted picture.

In 1965 Gorkov and Eliashberg \cite{ge} suggested a description of
level statistics in small disordered metal particles using the
random matrix theory (RMT) of Wigner-Dyson \cite{wigner,dyson}. At
first glance, the diagrammatic method of Ref. \cite{agd} had to
work for such a system but one could not see any indication on how
the formulae of RMT could be obtained diagrammatically. Of course,
the description of Ref. \cite{ge} was merely a hypothesis and the
RMT had not been used in the condensed matter before but nowadays
it looks rather strange that this problem did not attract an
attention.

The prediction of localization in thick wires for any disorder made by
Thouless \cite{thouless} could not be understood in terms of the traditional
summing of the diagrams either but, again, there was no attempt to clarify
this disagreement. Apparently, the diagrammatic methods were not very widely
used in that time and therefore not so many people were interested in
resolving such problems.

Actually, the discrepancies were not discussed in the literature
until 1979, the year when the celebrated work by Abrahams et al.
\cite{aalr} appeared. In this work, localization of all states for
any disorder already in $2D$ was predicted. This striking result
has attracted so much attention that it was simply unavoidable
that people started thinking about how to confirm it
diagrammatically. The only possibility could be that there were
some diverging quantum corrections to the classical conductivity
and soon the mechanism of such divergencies has been discovered
\cite{glk,aar,ar}.

\begin{figure}
\centerline{\psfig{file=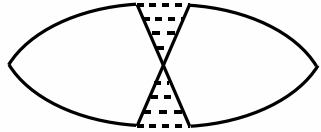,width=5.2cm}}
\caption{ Diverging contribution to conductivity (cooperon) }
\label{fig1}
\end{figure}

It turns out that the sum of a certain class of the diagrams with
intersecting impurity lines diverges in the limit of small frequencies $%
\omega \rightarrow 0$ in a low dimension $d\leq 2$. This happens
for any weak disorder and is a general phenomenon. The
corresponding contribution is represented in Fig. \ref{fig1}.

The ladder in this diagram can be considered as an effective mode
usually called now \textquotedblleft cooperon". This mode has a
form of the diffusion propagator and its contribution to the
conductivity $\sigma \left( \omega \right) $ can be
written in the form%
\begin{equation}
\sigma \left( \omega \right) =\sigma _{0}\left( 1-\frac{1}{\pi \nu }\int
\frac{1}{D_{0}\mathbf{k}^{2}-i\omega }\frac{d^{d}\mathbf{k}}{\left( 2\pi
\right) ^{d}}\right)   \label{a1}
\end{equation}
where $D_{0}=v_{0}^{2}\tau /3$ is the classical diffusion coefficient and $%
\sigma _{0}=2e^{2}\nu D_{0}$ is the classical conductivity. The parameters $%
v_{0}$ and $\nu $ are the Fermi velocity and density of states on the Fermi
surface.

Similar contributions arise also in other quantities. Eq.
(\ref{a1}) demonstrates that in the dimensions $d=0,1,2$ the
correction to conductivity diverges in the limit $\omega
\rightarrow 0$. It is very important that the dimension is
determined by the geometry of the sample. In this sense, small
disordered particles correspond to zero dimensionality, $d=0,$ and
wires to $d=1$.

The contribution coming from the diffusion mode, Eq. (\ref{a1}), is
conceptually very important because it demonstrates that the traditional
summation of the diagrams without the intersection of the impurity lines is
not necessarily applicable in low dimensionality. One can see that most
important contributions come from the diffusion modes that are obtained by
summation of infinite series of diagrams containing electron Green
functions.

The cooperon contribution, Eq. (\ref{a1}), has a simple physical meaning. It
is proportional to the probability for a scattered electron wave to come back
and interfere with itself \cite{khmelnitskii}. The interference implies the
quantum coherence and this condition is achieved at low temperatures. There
are many interesting effects related to this phenomenon but discussion of
these effects and experiments is beyond the scope of this chapter.

It is also relevant to mention that the cooperon contribution is
cut by an external magnetic field, which leads to a negative
magnetoresistance \cite{akll}. At the same time, higher order
contributions can still diverge in the limit $\omega \rightarrow
0$ and these divergencies are not avoidable provided the coherence
is not lost due to, e.g., inelastic processes.

In this way, one can reconcile the hypothesis about the
Wigner-Dyson level statistics in disordered metal particles and
assertion about the localization in thick wires and $2D$ films
with the perturbation theory in the disorder potential. The
divergences due to the contribution of the
diffusion modes make the perturbation theory inapplicable in the limit $%
\omega \rightarrow 0$ and therefore one does not obtain just the
classical conductivity using this approach. Of course, summing the
divergent quantum corrections is not sufficient to prove the
localization in the low dimensional systems and one should use
additional assumptions in order to confirm the statements.
Usually, the perturbation theory is supplemented by the scaling
hypothesis \cite{aalr} in order to make such far going
conclusions.

At the same time, the divergence of the quantum corrections to the
conductivity makes the direct analytical consideration very difficult for
small $\omega $ because even the summation of all orders of the perturbation
theory does not necessarily lead to the correct result. For example, the
formulae for the level-level correlation functions \cite{wigner, dyson}
contain oscillating parts that cannot be obtained in any order of the
perturbation theory.

All this meant that a better tool had to be invented for studying
the localization phenomena and quantum level statistics. Analyzing
the perturbation theory one could guess that a low energy theory
explicitly describing the diffusion modes rather than single
electrons might be an adequate method.

The first formulation of such a theory was proposed by Wegner \cite{wegner}
(actually, almost simultaneously with Ref. \cite{glk}). He expressed the
electron Green functions in terms of functional integrals over conventional
complex numbers $S\left( r\right) $, where $r$ is the coordinate, and
averaged over the disorder using the replica trick. Then, decoupling the
effective interaction by an auxiliary matrix field $Q$ he was able to
integrate over the field $S\left( r\right) $ and represent physical
quantities of interest in terms of a functional integral over the $N\times N$
matrices $Q,$ where $N$ is the number of replicas that had to be put to zero
at the end of the calculations. Assuming that the disorder is weak the
integral over the eigenvalues of the matrix $Q$ was calculated using the
saddle point approximation.

As a result, a field theory in a form of a so called $\sigma $-model was
obtained. Working with this model one has to integrate over $N\times N$
matrices $Q$ obeying the constraint $Q^{2}=1$. The $\sigma $-model is
renormalizable and renormalization group equations were written in Ref. \cite%
{wegner}. These equations agreed with the perturbation theory of Eq. (\ref%
{a1}) and with the scaling hypothesis of Ref. \cite{aalr}.

However, the saddle point approximation was not carefully worked out in\cite%
{wegner} because the saddle points were in the complex plane, while the
original integration had to be done over the real axis. This question was
addressed in the subsequent publications \cite{sw,elk}.

In the work \cite{sw}, the initial derivation of Ref. \cite{wegner} was done
more carefully shifting the contours of the integration into the complex
plane properly. In this way, one could reach the saddle point and integrate
over the eigenvalues of matrix $Q$ coming to the constraint $Q^{2}=1$. After
calculating this integral one is left with the integration over $Q$ that can
be written as
\begin{equation}
Q=U\Lambda U^{-1},\quad \Lambda =\left(
\begin{array}{cc}
1 & 0 \\
0 & -1%
\end{array}%
\right)   \label{a2}
\end{equation}%
where $U$ is an $2N\times 2N$ pseudo-orthogonal or pseudo-unitary
matrix. This matrices vary on a hyperboloid, which corresponds to
a noncompact group of the rotations. This group is quite unusual
for statistical physics.

In contrast, the method of \cite{elk} was based on representing
the electron Green functions in a form of functional integrals
over anticommuting Grassmann variables and the use of the replica
trick. One could average over the disorder as well and further
decouple the effective interaction by a gaussian integration over
$Q$. The integration over the anticommuting variables leads to an
integral over $Q$. The integral over the eigenvalues of $Q$ can be
calculated using, again, the saddle point method, while the
saddle points are now on the real axis. As a result, one comes to a $\sigma $%
-model with $Q$-fields of the form of Eq. (\ref{a2}). However, now one
obtains $2N\times 2N$ matrices $U$ varying on a sphere and the group of the
rotations is compact.

The difference in the symmetry groups of the matrices $Q$ of these two
approaches looked rather unusual and one could only hope that in the limit $%
N=0$ imposed by the replica method the results would have agree with each
other.

This is really so for the results obtained in Refs. \cite{sw,elk}
by using the renormalization group method or perturbation theory.
The compact replica $\sigma $-model of Ref. \cite{elk} has later
been extended by Finkelstein \cite{finkelstein}to interacting
electron systems. An additional topological term was added to this
model by Pruisken\cite{pruisken} for studying the
Integer Quantum Hall Effect. So, one could hope that the replica $\sigma $%
-models would help to solve many problems in the localization
theory.

However, everything turned out to be considerably more complicated
for non-perturbative calculations. Desperate attempts
\cite{efetovrep} to study the level-level statistics in a limited
volume and localization in disordered wires lead the present
author to the conclusion that the replica $\sigma $-model of Ref.
\cite{elk} could not give any reasonable formulae. Calculation of
the level-level correlation function using both the compact and
noncompact replica $\sigma $-models was discussed later by
Verbaarschot and Zirnbauer \cite{vz} with a similar result.
[Recently, formulae for several correlation functions for the
unitary ensemble ($\beta =2$) have nevertheless been obtained \cite%
{kanzieper} from the replica $\sigma $-models by viewing the
replica partition function as Toda Lattice and using links with
Panleve equations.]

The failure in performing non-perturbative calculations with the replica $%
\sigma $-models lead the present author to constructing another type of the $%
\sigma $-model that was not based on the replica trick. This
method was called supersymmetry method, although the word
\textquotedblleft supersymmetry" is often used in field theory in
a more narrow sense. The field theory derived for the disordered
systems using this approach has the same form of the $\sigma
$-model as the one obtained with the replica trick and all
perturbative calculations are similar \cite{efetov1}.

An attempt to calculate the level-level correlation function lead
to a real surprise: the method worked \cite{efetov2}  leading in a
rather simple way to the famous formulae for the level-level
correlation functions known in the Wigner-Dyson theory
\cite{wigner,dyson}, thus establishing the relevance of the latter
to the disordered systems. Since then one could use the RMT for
calculations of various physical quantities in mesoscopic systems
or calculate directly using the zero-dimensional supermatrix
$\sigma $-model.

The calculation of the level correlations in small disordered
systems
followed by the full solution of the localization problem in wires \cite%
{efetov3}, on the Bethe lattice and in high dimensionality
\cite{efetov84,zirnbauerbethe,efetov87,efetov87a,efetov88}. After
that it has become clear that the supersymmetry technique is
really an efficient tool suitable for solving various problems of
theory of disordered metals.

By now several reviews and a book have been published \cite%
{efetovreview,vwz,book,gmw,mirlin,evers} where numerous problems
of disordered, mesoscopic and ballistic chaotic system are
considered and solved using the supersymmetry method. The
interested reader can find all necessary references in those
publications.

The present paper is not a complete review of all the works done
using the supersymmetry method. Instead, I describe here the main
steps leading to the supermatrix $\sigma $-model and first
problems solved using this approach. I will try to summarize at
the end what has become clear in the last almost $30 $ years of
the development and what problems await their resolution.

\section{Supermatrix non-linear $\protect\sigma $-model.}

The supersymmetry method is based on using both integrals over
conventional complex numbers $S_{i}$ and anticommuting Grassmann
variables $\chi _{i}$
obeying the anticommutation relations%
\begin{equation}
\chi _{i}\chi _{j}+\chi _{j}\chi _{i}=0  \label{a3a}
\end{equation}%
The integrals over the Grassmann variables are used following the
definition
given by Berezin \cite{berezin}%
\begin{equation}
\int d\chi _{i}=0,\quad \int \chi _{i}d\chi _{i}=1  \label{a3}
\end{equation}%
With this definition one can write the Gaussian integral $I_{A}$
over the
Grassmann variables as%
\begin{equation}
I_{A}=\int \exp \left( -\chi ^{+}A\chi \right)
\prod_{i=1}^{N}d\chi _{i}^{\ast }d\chi _{i}=\det A,  \label{a4}
\end{equation}%
which is different from the corresponding integral over complex
numbers by presence of $\det A$ instead of $\left( \det A\right)
^{-1}$ in the R.H.S. In Eq. (\ref{a4}), $\chi $ is a vector having
as components the anticommuting variables $\chi _{i}$ ($\chi ^{+}$
is its transpose with components $\chi ^{\ast }$) and $A$ is an
$N\times N$ matrix.

One can introduce supervectors $\Phi $ with the components $\Phi
_{i}$,
\begin{equation}
\Phi _{i}=\left(
\begin{array}{c}
\chi _{i} \\
S_{i}%
\end{array}%
\right)   \label{a5}
\end{equation}%
and write gaussian integrals for these quantities%
\begin{equation}
I_{S}=\pi ^{-N}\int \exp \left( -\Phi ^{+}F\Phi \right)
\prod_{i}^{N}d\chi _{i}^{\ast }d\chi _{i}dS^{\ast }dS=SDet\,F
\label{a6}
\end{equation}%
In Eq. (\ref{a6}), $F$ is a supermatrix with block elements of the form%
\begin{equation}
F_{ik}=\left(
\begin{array}{cc}
a_{ik} & \sigma _{ik} \\
\rho _{ik} & b_{ik}%
\end{array}%
\right)   \label{a7}
\end{equation}%
where $a_{ik}$ and $b_{ik}$ are complex numbers and $\sigma
_{ik},\rho _{ik}$
are Grassmann variables. The superdeterminant (Berezinian) $SDet\,F$ in Eq. (%
\ref{a6}) has the form
\begin{equation}
SDet\,F=\det \left( a-\sigma b^{-1}\rho \right) \det b^{-1}
\label{a8}
\end{equation}

Another important operation is supertrace $STr$%
\begin{equation}
STrF=Tra-Trb  \label{a9}
\end{equation}%
Using these definitions one can operate with supermatrices in the
same way
as with conventional matrices. Note a very important consequence of Eq. (\ref%
{a6}) for supermatrices $F_{0}$ that do not contain the
anticommuting
variables and are equal to unity in the superblocks $F_{ik}$ in Eq. (\ref{a7}%
) ( $a_{ik}=b_{ik}$). In this case one obtains%
\begin{equation}
I_{S}\left[ F_{0}\right] =1  \label{a10}
\end{equation}%
For such supermatrices one can write a relation that is the basis
of the
supersymmetry method in disordered metals%
\begin{equation}
F_{0ik}^{-1}=\int \Phi _{i}\Phi _{k}^{+}\exp \left( -\Phi
^{+}F\Phi \right) d\Phi   \label{a11}
\end{equation}%
where $d\Phi =$   $\pi ^{-N}\prod_{i}^{N}d\chi _{i}^{\ast }d\chi
_{i}dS^{\ast }dS$.

The weight denominator in the integral in Eq. (\ref{a11}) is
absent and this form is analogous to what one has using the
replica trick. Applying this representation to correlation
functions describing disordered systems one can average over the
disorder just in the beginning before making
approximations. This is what is done when deriving the supermatrix $\sigma $%
-model and let me sketch this derivation.

Many quantities of interest can be expressed in terms of products
of retarded $G_{\varepsilon }^{R}$ and advanced $G_{\varepsilon
}^{A}$ Green functions of the Schrodinger equation. Using Eq.
(\ref{a11}) one can write
these functions as integrals over supervectors $\Phi $ (see \cite%
{efetovreview,book})%
\begin{eqnarray}
G_{\varepsilon }^{R,A}\left( y,y^{\prime }\right)  &=&\mp i\int
\Phi _{\alpha }\left( y\right) \Phi _{\alpha }^{+}\left( y^{\prime
}\right)   \label{a12}
\\
&&\times \exp \left[ i\int \Phi ^{+}\left( x\right) \left( \pm
\left( \varepsilon -H\right) +i\delta \right) \Phi \left( x\right)
dx\right] D\Phi ^{+}D\Phi   \notag
\end{eqnarray}%
where $x$ and $y$ stand for both the space and spin variables.

The Hamiltonian $H$ in Eq. (\ref{a12})consists of the regular
$H_{0}$ and random $H_{1}$
parts%
\begin{equation}
H=H_{0}+H_{1},\;\left\langle H_{1}\right\rangle =0  \label{a13}
\end{equation}%
where the angular brackets $\left\langle ...\right\rangle $ stand
for the averaging over the disorder.

The most important contribution to such quantities as conductivity
and density-density correlation function is expressed in terms of
a product
\begin{equation}
K_{\omega }\left( \mathbf{r}\right) =2\left\langle G_{\varepsilon
-\omega
}^{A}\left( \mathbf{r,}0\right) G_{\varepsilon }^{R}\left( 0,\mathbf{r}%
\right) \right\rangle   \label{a14}
\end{equation}%
where $\mathbf{r}$ is a coordinate and $\omega $ is the frequency
of the external electric field.

In order to express the function $K_{\omega }\left(
\mathbf{r}\right) $ in terms of an integral over supervectors one
should double the size of the supervectors. Introducing such
supervectors $\psi $ one represents the function $K_{\omega
}\left( \mathbf{r}\right) $ in terms of a gaussian integral
without a weight denominator. This allows one to average
immediately this function over the random part. In the case of
impurities described by a white noise disorder potential $u\left(
\mathbf{r}\right) $ one comes after averaging to the following
expression
\begin{equation}
K_{\omega }\left( \mathbf{r}\right) =2\int \psi _{\alpha
}^{1}\left( 0\right)
\psi _{\alpha }^{1}\left( \mathbf{r}\right) \psi _{\beta }^{2}\left( \mathbf{%
r}\right) \psi _{\beta }^{2}\left( 0\right) \exp \left( -L\right)
D\psi \label{a15}
\end{equation}%
where
\begin{equation}
L=\int \left[ i\bar{\psi}\left( \varepsilon -H_{0}\right) \psi +\frac{1}{%
4\pi \nu \tau }\left( \bar{\psi}\psi \right) ^{2}-\frac{i\left(
\omega +i\delta \right) }{2}\bar{\psi}\Lambda \psi \right]
d\mathbf{r}  \label{a16}
\end{equation}%
Eq. (\ref{a16}) was obtained assuming the averages%
\begin{equation}
\left\langle u\left( \mathbf{r}\right) u\left( \mathbf{r}^{\prime
}\right) \right\rangle =\frac{1}{2\pi \nu \tau }\delta \left(
\mathbf{r-r}^{\prime }\right) ,\quad \left\langle u\left(
\mathbf{r}\right) \right\rangle =0 \label{a17}
\end{equation}%
where $\nu $ is the density of states and $\tau $ is the elastic
scattering time.

The fields $\bar{\psi}$ in Eqs. (\ref{a15}, \ref{a16}) are conjugate to $%
\psi ,$ the matrix $\Lambda $ is in the space of the
retarded-advanced Green functions and equals
\begin{equation}
\Lambda =\left(
\begin{array}{cc}
1 & 0 \\
0 & -1%
\end{array}%
\right)   \label{a18}
\end{equation}
The infinitesimal $\delta \rightarrow +0$ is added to guarantee
the convergence of the integrals over the commuting components $S$
of the supervectors $\psi $.

The Lagrangian $L$, Eq. (\ref{a16}) has a form corresponding to a
field theory of interacting particles. Of course, physically this
interaction is fictitious but this formal analogy helps one to use
approximations standard for many body theories.

The first approximation done in the supersymmetry method is
singling out slowly varying
pairs in the interaction term. This is done writing it as%
\begin{eqnarray}
L_{int} &=&\frac{1}{4\pi \nu \tau }\int \left( \bar{\psi}\psi \right) ^{2}d%
\mathbf{r=}\frac{1}{4\pi \nu \tau }\sum_{\mathbf{p}_{1}+\mathbf{p}_{2}+%
\mathbf{p}_{3}+\mathbf{p}_{4}=0}\left( \bar{\psi}_{\mathbf{p}_{1}}\psi _{%
\mathbf{p}_{2}}\right) \left( \bar{\psi}_{\mathbf{p}_{3}}\psi _{\mathbf{p}%
_{4}}\right)  \notag \\
&\thickapprox &\frac{1}{4\pi \nu \tau
}\sum_{p_{1},p_{2},q<q_{0}}[\left( \bar{\psi}_{\mathbf{p}_{1}}\psi
_{-\mathbf{p}_{1}+\mathbf{q}}\right) \left(
\bar{\psi}_{\mathbf{p}_{2}}\psi
_{-\mathbf{p}_{2}-\mathbf{q}}\right)  \notag
\\
&&+\left( \bar{\psi}_{\mathbf{p}_{1}}\psi _{\mathbf{p}_{2}}\right)
\left(
\bar{\psi}_{-\mathbf{p}_{2}-\mathbf{q}}\psi _{-\mathbf{p}_{1}+\mathbf{q}%
}\right) +\left( \bar{\psi}_{\mathbf{p}_{1}}\psi
_{\mathbf{p}_{2}}\right)
\left( \bar{\psi}_{-\mathbf{p}_{1}+\mathbf{q}}\psi _{-\mathbf{p}_{2}-\mathbf{%
q}}\right) ]  \label{a19}
\end{eqnarray}%
where $q_{0}$ is a cutoff parameter, $q_{0}<1/l$, where $l$ is the
mean free path.

The next step is making a Hubbard-Stratonovich transformation
decoupling the products of slowly varying pairs by auxiliary
slowly varying fields. The term in the second line in Eq.
(\ref{a19}) is not important and the terms in the third line are
equal to each other provided one uses the form of the supervectors
$\psi $ of Refs. \cite{efetovreview,book}.

After the decoupling one obtains an effective Lagrangian quadratic
in the
fields $\psi ,\bar{\psi}$ and one can integrate out the fields $\psi ,\bar{%
\psi}$ in Eq. (\ref{a15}) and obtain a functional integral over
the supermatrix field $Q\left( \mathbf{r}\right) .$ The
corresponding free energy functional $F\left[ Q\right] $ takes the
form
\begin{equation}
F\left[ Q\right] =\int \left[ -\frac{1}{2}STr\ln \left( \varepsilon -H_{0}-%
\frac{\left( \omega +i\delta \right) }{2}\Lambda -\frac{iQ\left( \mathbf{r}%
\right) }{2\tau }\right) +\frac{\pi \nu }{8\tau }STrQ^{2}\right]
dr \label{a20}
\end{equation}%
and physical quantities should be obtained integrating correlation
functions containing $Q$ over $Q$ with the weight $\exp \left(
-F\left[ Q\right] \right) $.

The integrals with $F\left[ Q\right] $ can be simplified using
the saddle point approximation. The position of the minimum of $F\left[ Q%
\right] $ is found in the limit $\omega \rightarrow 0$ by solving
the
equation%
\begin{equation}
Q=\frac{i}{\pi \nu }\left[ \left( H_{0}+\frac{i}{2\tau }Q\left( \mathbf{r}%
\right) \right) ^{-1}\right] _{\mathbf{r,r}}
\end{equation} \label{a21}

One can find rather easily a coordinate independent solution of Eq. (\ref{a21}%
). Writing $H_{0}$ in a general form as
\begin{equation}
H_{0}=\varepsilon \left( -i\nabla _{r}\right) -\varepsilon _{0}
\label{a22}
\end{equation}%
and Fourier transforming the latter, one should calculate the
integral over the momenta $\mathbf{p}$. In the limit $\varepsilon
_{0}\tau \gg 1$ one comes to the general solution%
\begin{equation}
Q^{2}=1  \label{a23}
\end{equation}%
Although the supermatrix $Q^{2}$ is fixed by Eq. (\ref{a23}), the
supermatrix $Q$ is not. Supermatrices $Q$ of the form of Eq.
(\ref{a2}) are
solutions for any $8\times 8$ supermatrices $U$ satisfying the condition $U%
\bar{U}=1$. With this constraint they are neither unitary nor
pseudo-unitary as it was in Refs. \cite{sw,elk}. Actually, they
consist of both unitary and pseudo-unitary sectors
\textquotedblleft glued" by the anticommuting variables. This
unique symmetry is extremely important for basic properties of
many physical quantities.

The degeneracy of the minimum of the free energy functional
$F\left[ Q\right] $ results in the existence of gapless in the
limit $\omega \rightarrow 0$ excitations (Goldstone modes). This
are diffusion modes: so called \textquotedblleft cooperons" and
\textquotedblleft diffusons". These modes formally originate from
fluctuating $Q$ obeying the constraint (\ref{a23}).

In order to write the free energy functional describing the
fluctuations we
assume that supermatrices $Q\left( \mathbf{r}\right) $ obeying Eq. (\ref{a23}%
) slowly vary in space. Assuming that $\omega $ is small, $\omega \tau \ll 1$%
, but finite and expanding $F\left[ Q\right] $ in this quantity
and
gradients of $Q$ one comes to the supermatrix $\sigma $-model%
\begin{equation}
F\left[ Q\right] =\frac{\pi \nu }{8}\int STr\left[ D_{0}\left(
\nabla Q\right) ^{2}+2i\left( \omega +i\delta \right) \Lambda
Q\right] d\mathbf{r} \label{a24}
\end{equation}%
where $D_{0}=v_{0}^{2}\tau /d$ is the classical diffusion
coefficient ($v_{0}$
is the Fermi velocity and $d$ is the dimensionality of the sample) and the $%
8\times 8$ supermatrix $Q$ obeys the constraint (\ref{a23}).

Calculation of, e.g., the function $K_{\omega }\left(
\mathbf{r}\right) $,
Eq. (\ref{a15}), reduces to calculation of a functional integral over $Q$%
\begin{equation}
K_{\omega }\left( \mathbf{r}\right) =2\int Q_{\alpha \beta
}^{12}\left( 0\right) Q_{\beta \alpha }^{21}\left(
\mathbf{r}\right) \exp \left( -F\left[ Q\right] \right) DQ
\label{a25}
\end{equation}%

Eqs. (\ref{a24}, \ref{a25}) is a reformulation of the initial
problem of disordered metal in terms of a field theory that does
not contain disorder because the averaging over the initial
disorder has already been carried out.
The latter enters the theory through the classical diffusion coefficient $%
D_{0}$. The supermatrix $\sigma $-model, described by Eq.
(\ref{a24}) resembles $\sigma $-models used for calculating
contributions of spin waves for magnetic materials. At the same
time, the noncompactness of the symmetry group of the
supermatrices $Q$ makes this $\sigma $-model unique.

In order to obtain classical formulae and first quantum
corrections one can parametrize the supermatrix $Q$ as
\begin{equation}
Q=W+\Lambda \left( 1-W^{2}\right) ^{1/2},\text{\quad }W=\left(
\begin{array}{cc}
0 & Q^{12} \\
Q^{21} & 0%
\end{array}%
\right)   \label{a25a}
\end{equation}%
and make an expansion in $W$ in Eqs. (\ref{a24}, \ref{a25}).
Keeping quadratic in $W$ terms both in $F\left[ Q\right] $ and in
the pre-exponential in Eq. (\ref{a25}) one has to compute Gaussian
integrals over $W$. Fourier transforming the function $K_{\omega
}$, one obtains
\begin{equation}
K_{\omega }\left( \mathbf{k}\right) =\frac{4\pi \nu }{D_{0}\mathbf{k}%
^{2}-i\omega }  \label{a25b}
\end{equation}%
Eq. (\ref{a25b}) is the classical diffusion propagator. Taking
into account higher orders in $W$ one can compute weak
localization corrections to the
diffusion coefficient. The first order correction is written in Eq. (\ref{a1}%
).

The precise symmetry of $Q$ depends on the presence of magnetic or
spin-orbit interactions. In analogy with symmetries of random
matrix ensembles in the Wigner-Dyson theory \cite{wigner,dyson}
one distinguishes between the orthogonal  ensemble (both magnetic
and spin orbit interactions are absent), unitary (magnetic
interactions are present) and symplectic (spin-orbit interactions
are present but magnetic interactions are absent).

Actually more symmetry classes are possible. They are fully
classified by Altland and Zirnbauer \cite{az}.

In the next sections solutions of several important problems
solved with the help of the $ \sigma $-model, Eq. (\ref{a24}),
will be presented.

\section{Level statistics in small metal particles. }

The first non-trivial problem solved with the supermatrix $\sigma
$-model was the problem of describing the level statistics in
small disordered metal particles. At first glance, this problem is
not related to the Anderson localization. However, in the language
of the $\sigma $-model the solutions of these problems is study of
the field theory, Eq. (\ref{a24}), in different dimensions. The
localization can be obtained in the dimensions $d=1,2$ and $3,$
while the Wigner-Dyson level statistics can be obtained for the
zero dimensional version of the $\sigma $-model.

What is the zero dimensionality of the free energy functional $F\left[ Q%
\right] $, Eq. (\ref{a24}), can easily be understood. In a finite
volume the space harmonics are quantized. The lowest harmonics
corresponds to the homogeneous in the space supermatrix $Q$. The
energy of the first excited harmonics $E_{1}$ can be estimated as
\begin{equation}
E_{1}=E_{c}/\Delta   \label{b1}
\end{equation}%
where energy $E_{c}$,
\begin{equation}
E_{c}=\pi ^{2}D_{0}/L^{2}  \label{b2}
\end{equation}%
is usually called the Thouless energy.

The other energy scale $\Delta $,%
\begin{equation}
\Delta =\left(\nu V\right) ^{-1},  \label{b3}
\end{equation}
where $V$ is the volume, is the mean level spacing.

It is clear from Eqs. (\ref{a24}, \ref{a25}) that in the limit
\begin{equation}
E_{c}\gg \Delta ,\,\omega \text{ }  \label{b4}
\end{equation}%
one may keep in these equations only the zero space harmonics of
$Q$, so that this supermatrix does not depend on the coordinates.
One can interpret this limit as zero dimensional one and replace
the functional $F\left[ Q\right]
$ by the function $F_{0}\left[ Q\right] $,%
\begin{equation}
F_{0}\left[ Q\right] =\frac{i\pi \left( \omega +i\delta \right) }{4\Delta }%
STr\left( \Lambda Q\right)   \label{b5}
\end{equation}

The function $R\left( \omega \right) $ that determines the
correlation between the energy levels is introduced as
\begin{equation}
R\left( \omega \right) =\left\langle \frac{\Delta ^{2}}{\omega }%
\sum_{k,m}\left( n\left( \varepsilon _{k}\right) -n\left(
\varepsilon _{m}\right) \right) \delta \left( \omega -\varepsilon
_{m}+\varepsilon _{k}\right) \right\rangle   \label{b6}
\end{equation}%
It is proportional to the probability of finding two levels at a distance $%
\omega $.

Using the supersymmetry approach one can represent the functions
$R\left(
\omega \right) $ in terms of a definite integral over the supermatrices $Q$%
\begin{equation}
R\left( \omega \right) =\frac{1}{2}-\frac{1}{2}Re\int
Q_{11}^{11}Q_{11}^{22}\exp \left( -F_{0}\left[ Q\right] \right) dQ
\label{b7}
\end{equation}%
In order to calculate the integral in Eq. (\ref{b7}) one should
choose a certain parametrization for the supermatrix $Q$.

It is convenient to write the supermatrix $Q$ in the form
\begin{equation}
Q=UQ_{0}\bar{U},\quad Q_{0}=\left(
\begin{array}{cc}
\cos \hat{\theta} & i\sin \hat{\theta} \\
-i\sin \hat{\theta} & -\cos \hat{\theta}%
\end{array}%
\right) ,\quad U=\left(
\begin{array}{cc}
u & 0 \\
0 & v%
\end{array}%
\right)   \label{b8}
\end{equation}%
where all anticommuting variables are packed in the supermatrix
blocks $u$ and $v$. It is clear that the (pseudo) unitary
supermatrix $U$ commutes with $\Lambda $, which drastically
simplifies the integrand in Eq. (\ref{b7}).

Instead of the integration in Eq. (\ref{b7}) over the elements of
the supermatrix $Q$ with the constraint (\ref{a23}) one can
integrate over the elements of the matrix $\hat{\theta}$ and the
matrices $u$ and $v$. Of course, it is necessary to write a proper
Jacobian (Berezinian) of the transformation to these variables.
The latter depends only on the elements of $\hat{\theta}$ and
therefore the elements of $u$ and $v$ appear only in the
pre-exponential in Eq. (\ref{b7}). The integration over the
supermatrices $u$ and $v$ is quite simple and one comes to
definite integrals over the elements of $\hat{\theta}$.

The number of the independent variables in the blocks
$\hat{\theta}$ depends on the ensemble considered. The
supermatrices $Q$ written for the unitary ensemble have the
simplest structure and the blocks $\hat{\theta}$ contains only $2$
variables $0<\theta <\pi $ and $0<\theta _{1}<\infty $. The
corresponding blocks $\hat{\theta}$ for the orthogonal and
symplectic ensembles contain $3$ independent variables. All the
transformations are described in details in Refs.
\cite{efetovreview,vwz,book}.

In order to get an idea about what one obtains after the
integration over $u$ and $v$ in Eq. (\ref{b8}), I write here an
expression for the unitary
ensemble only%
\begin{equation}
R\left( \omega \right) =1+\frac{1}{2}Re\int_{1}^{\infty
}\int_{-1}^{1}\exp \left[ i\left( x+i\delta \right) \left( \lambda
_{1}-\lambda \right) \right] d\lambda _{1}d\lambda   \label{b9}
\end{equation}%
where $x=\pi \omega /\Delta $, $\lambda _{1}=\cosh \theta _{1}$, and $%
\lambda =\cos \theta $.

So, the calculation of the level-level correlation function is
reduced to an integral over $2$ or $3$ variables depending on the
ensemble considered. The
final result for the orthogonal $R_{orth}\left( \omega \right) $, unitary $%
R_{unit}\left( \omega \right) $, and symplectic $R_{sympl}\left(
\omega \right) $ ensembles calculated using Eq. (\ref{b7}) takes
the following form
\begin{equation}
R_{orth}\left( \omega \right) =1-\frac{\sin
^{2}x}{x^{2}}-\frac{d}{dx}\left( \frac{\sin x}{x}\right)
\int_{1}^{\infty }\frac{\sin xt}{t}dt  \label{b10}
\end{equation}%
\begin{equation}
R_{unit}\left( \omega \right) =1-\frac{\sin ^{2}x}{x^{2}}
\label{b11}
\end{equation}%
\begin{equation}
R_{sympl}\left( \omega \right) =1-\frac{\sin ^{2}x}{x^{2}}+\frac{d}{dx}%
\left( \frac{\sin x}{x}\right) \int_{0}^{1}\frac{\sin xt}{t}dt
\label{b12}
\end{equation}%
Eqs. (\ref{b10}-\ref{b12}) first obtained for the disordered metal
particles \cite{efetov2,efetovreview} identically agree with the
corresponding formulae of the Wigner-Dyson theory
\cite{wigner,dyson} obtained from the ensembles of random
matrices. This agreement justified the application of the RMT for
small disordered particles suggested in Ref. \cite{ge}.

Actually, to the best of my knowledge, this was the first explicit
demonstration that RMT could correspond to a real physical system.
Its original application to nuclear physics was in that time
phenomenological and confirmed by neither analytical nor numerical
calculations.

A direct derivation of Eqs. (\ref{b10}-\ref{b12}) from gaussian
ensembles of the random matrices using the supermatrix approach
was done in the review \cite{vwz}. This allowed the authors to
compute certain average compound-nucleus cross sections that could
not be calculated using the standard RMT route.

The proof of the applicability of the RMT to the disordered
systems was followed by the conjecture of Bohigas, Giannonni and
Schmid \cite{bgs} about the possibility of describing by RMT the
level statistics in classically chaotic clean billiards.
Combination of the results for clean and disordered small systems
(billiards) has established the validity of the use of RMT in
mesoscopic systems. Some researches use for explicit calculations
methods of RMT but many others use the supermatrix
zero-dimensional $\sigma $-model
(for review see, e.g. \cite{gmw,suptrace,been}). At the same time, the $%
\sigma $-model is applicable to a broader class of systems than
the Wigner-Dyson RMT because it can be used in higher dimensions
as well. Actually, one can easily go beyond the zero
dimensionality taking higher space harmonics in $F\left[ Q\right]
$, Eq. (\ref{a24}). In this case, the universality of Eqs.
(\ref{b10}-\ref{b11}) is violated. One can study this limit for
$\omega \gg \Delta $ using also the standard diagrammatic
expansions of Ref. \cite{agd} and this was done in Ref.
\cite{altshk}.

The other versions of the $\sigma $-model (based on the replica
trick and Keldysh Green functions) have not shown a comparable
efficiency for studying
the mesoscopic systems, although the formula for the unitary ensemble, Eq. (%
\ref{b11}), has been obtained by these approaches \cite{kanzieper,
ak}.

The results reviewed in this section demonstrate that the
development of the theory of the energy level statistics in small
systems and of related phenomena in mesoscopic systems have been
tremendously influenced by the ideas of the Anderson localization
because important results have been obtained by methods developed
for studying the latter.

\section{Anderson localization in quantum wires \label{wires}}

The one dimensional $\sigma $-model corresponds to quantum wires.
These objects are long samples with a finite cross-section $S$
that should be sufficiently large,
\begin{equation}
Sp_{0}^{2}\gg 1,  \label{c0}
\end{equation}%
where $p_{0}$ is the Fermi momentum. In other words, the number of
transversal channels should be large. This condition allows one to
neglect non-homogeneous in the transversal direction variations of
$Q$. Of course, the inequality $\varepsilon _{0}\tau \gg 1$ should
be fulfilled as before.

Then, the $\sigma $-model can be written in the form
\begin{equation}
F\left[ Q\right] =\frac{\pi \tilde{\nu}}{8}\int \left[ D_{0}\left( \frac{dQ}{%
dx}\right) ^{2}+2i\omega \Lambda Q\right] dx,  \label{c1}
\end{equation}%
where $\tilde{\nu}=\nu S$.

Again, depending of the presence of magnetic and/or spin-orbit
interactions the model has different symmetries (orthogonal,
unitary and symplectic). It is important to emphasize that Eq.
(\ref{c1}) is not applicable for disordered chains or thin wires
where the inequality (\ref{c0}) is not fulfilled. However, the
explicit solutions show that the low frequency behavior of all
these systems is the same.

Computation of the correlation function $K_{\omega }\left( x\right) $, Eq. (%
\ref{a25}), with the one-dimensional $\sigma $-model can be
performed using the transfer matrix technique. Following this
method one reduces the calculation of the functional integral in
Eq. (\ref{a25}) to solving an effective Schr\"{o}dinger equation
in the space of the elements of the supermatrix $Q$ and
calculating matrix elements of $Q$ entering the pre-exponential in
Eq. (\ref{a25}). This has been done in Ref. \cite{el} and
presented also in the subsequent publications
\cite{efetovreview,book}.

At first glance, this procedure looks very complicated due to a
large number of the elements in the supermatrices $Q.$
Fortunately, the symmetries of the free energy functional $F\left[
Q\right] $ in Eq. (\ref{c1}) help one again to simplify the
calculations.

In order to derive the transfer matrix equations one should
subdivide the wire into small slices and write recursive equations
taking at the end the continuous limit. Instead of this artificial
subdivision it is more instructive to consider a realistic model
of a chain of grains coupled by tunnelling. The free energy
functional $F_{J}\left[ Q\right] $ for such a
chain can be written in the form%
\begin{equation}
F_{J}\left[ Q\right] =STr\left(
-\sum_{i,j}J_{ij}Q_{i}Q_{j}+\frac{i\left( \omega +i\delta \right)
\pi }{4\Delta }\sum_{i}\Lambda Q_{i}\right) \label{c2}
\end{equation}
where $J_{ij}=J$ for nearest neighbors and $J_{ij}=0$ otherwise.
The summation runs in Eq. (\ref{c2}) over the grains. The coupling
constant $J$ can be expressed in terms of the matrix elements of
the tunnelling from grain to grain $T_{ij}$ but at the moment this
explicit relation is not important.

In the limit $J\gg 1$, only small variations of the supermatrix
$Q$ in space are important and the functional $F_{J}\left[
Q\right] ,$ Eq. (\ref{c2}), can be approximated by $F\left[
Q\right] $, Eq. (\ref{c1}). The classical diffusion coefficient
$D_{0}$ corresponding to Eq. (\ref{c2}) takes the form
\begin{equation}
D_{0}=\frac{4\Delta }{\pi }\sum_{i}J_{ij}\left( r_{i}-r_{j}\right)
^{2} \label{c3}
\end{equation}

The correlation function $K_{\omega }$, Eq. (\ref{a25}), should
also be taken at the discrete coordinates $r_{i}$ numerating the
grains. Then, it
can be re-written identically in the form%
\begin{eqnarray}
K_{\omega }\left( r_{1},r_{2}\right)  &=&2\pi ^{2}\nu \tilde{\nu}%
\int \Psi \left( Q_{1}\right) \left( Q_{1}\right) _{11}^{12}  \label{c4} \\
&&\times \Gamma \left( r_{1},r_{2};Q_{1},Q_{2}\right) \left(
Q_{2}\right) _{11}^{21}\Psi \left( Q_{2}\right) dQ_{1}dQ_{2}
\notag
\end{eqnarray}%
where the kernel $\Gamma \left( r_{1},r_{2};Q_{1},Q_{2}\right) $
is the partition function of the segment between the points
$r_{1}$ and $r_{2}$. It is assumed that integration for this
kernel is performed over all $Q$ except $Q_{1}$ and $Q_{2}$ at the
points $r_{1}$ and $r_{2}$. So the kernel $\Gamma
\left( r_{1},r_{2};Q_{1},Q_{2}\right) $ depends on supermatrices $Q_{1}$, $%
Q_{2}$ and distances $r_{2}-r_{1}$ (the point $r_{2}$ is to the
right of the point $r_{1}$). The function $\Psi \left( Q\right) $
is the partition function of the parts of the wire located to the
right of the point $r_{2}$ and to the left of the point $r_{1}$.
This function depends only on the supermatrix $Q$ at the end
points $r_{1}$ or $r_{2}$.

Comparing the functions $\Psi \left( Q\right) $ at neighboring
grains one
comes to the following equation%
\begin{equation}
\Psi \left( Q\right) =\int N\left( Q,Q^{\prime }\right)
Z_{0}\left( Q^{\prime }\right) \Psi \left( Q^{\prime }\right)
dQ^{\prime }  \label{c5}
\end{equation}%
where
\begin{eqnarray}
N\left( Q,Q^{\prime }\right)  &=&\exp \left( \frac{\alpha
}{4}STrQQ^{\prime
}\right) ,\quad \alpha =8J  \label{c6} \\
Z_{0}\left( Q\right)  &=&\exp \left( \frac{\beta }{4}STr\Lambda
Q\right) ,\quad \beta =\frac{-i\left( \omega +i\delta \right) \pi
}{\Delta }  \notag
\end{eqnarray}%
A similar equation can be written for the kernel $\Gamma \left(
r_{1},r_{2};Q_{1},Q_{2}\right) .$ Comparing this function at the
neighboring points $r$ and $r+1$ one obtains the recurrence equation%
\begin{eqnarray}
&&\Gamma \left( r,r^{\prime };Q,Q^{\prime }\right) -\int N\left(
Q,Q^{\prime \prime }\right) Z_{0}\left( Q^{\prime \prime }\right)
\Gamma \left( r+1,r^{\prime };Q^{\prime \prime },Q^{\prime
}\right) dQ^{\prime \prime }
\notag \\
&=&\delta _{rr^{\prime }}\delta \left( Q-Q^{\prime }\right)
\label{c7a}
\end{eqnarray}%
The $\delta $-function entering Eq. (\ref{c7a}) satisfies the
usual equality
\begin{equation}
\int f\left( Q^{\prime }\right) \delta \left( Q-Q^{\prime }\right)
dQ^{\prime }=f\left( Q\right)   \label{c8}
\end{equation}%
Eqs. (\ref{c4}-\ref{c8}) reduce the problem of calculation of a
functional integral over $Q\left( r\right) $ to solving the
integral equations and calculation of the integrals with their
solutions. In the limit $J\gg 1$ the integral equations can be
reduced to differential ones. Their solution can be sought using
again the parametrization (\ref{b8}). The function $\Psi
\left( Q\right) $ is assumed to be a function of the elements of the block $%
\hat{\theta}$. Then, one obtains the differential equation for
$\Psi $ in
the form%
\begin{equation}
\mathcal{H}_{0}\Psi =0  \label{c9}
\end{equation}%
The explicit form of the operator $\mathcal{H}_{0}$ depends on the
ensemble considered. The simplest equation is obtained for the
unitary ensemble for
which the operator $\mathcal{H}_{0}$ takes the form%
\begin{equation}
\mathcal{H}_{0}=-\frac{1}{2\pi \tilde{\nu}D_{0}}\left[ \frac{1}{J_{\lambda }}%
\frac{\partial }{\partial \lambda }J_{\lambda }\frac{\partial
}{\partial
\lambda }+\frac{1}{J_{\lambda }}\frac{\partial }{\partial \lambda _{1}}%
J_{\lambda }\frac{\partial }{\partial \lambda _{1}}\right]
-i\left( \omega +i\delta \right) \pi \tilde{\nu}\left( \lambda
_{1}-\lambda \right) \label{c10}
\end{equation}%
where
\begin{equation*}
J_{\lambda }=\left( \lambda _{1}-\lambda \right) ^{-2}
\end{equation*}

Similar equations can be written for the central part entering Eq.
(\ref{c4}).

Solving these equations and substituting the solutions into Eq.
(\ref{c4}) one can determine (at least numerically) the frequency
dependence of the function $K_{\omega }\left( r_{1},r_{2}\right) $
and, hence, of the conductivity for all frequencies in the region
$\omega \tau \ll 1$ and distances $\left\vert
r_{1}-r_{2}\right\vert p_{0}\gg 1$.

The calculation becomes considerably simpler in the most
interesting case of low frequencies $\omega \ll \left(
\tilde{\nu}^{2}D_{0}\right) ^{-1}$. In this limit, the main
contribution into the integral in Eq. (\ref{c4}) comes from large
$\lambda _{1}\gg 1$ and the solution $\Psi $ of Eq. (\ref{c9}) is
a function of only this variable.

Introducing a new variable
\begin{equation}
z=-i\omega 2\pi ^{2}\tilde{\nu}^{2}D_{0}\lambda _{1}  \label{c11}
\end{equation}%
one can reduce Eq. (\ref{c10}) to the form%
\begin{equation}
-z\frac{d^{2}\Psi \left( z\right) }{dz^{2}}+\Psi \left( z\right)
=0 \label{c12}
\end{equation}%
with the boundary condition
\begin{equation}
\Psi \left( 0\right) =1  \label{c12a}
\end{equation}

The Fourier transformed function $K_{\omega }\left( k\right) $ takes the form%
\begin{equation}
K_{\omega }\left( k\right) =\frac{4\pi \nu A\left( k\right) }{-i\omega }%
,\qquad A\left( k\right) =\int_{0}^{\infty }\left( \Phi _{k}\left(
z\right) +\Phi _{-k}\left( z\right) \right) \Psi \left( z\right)
dz,  \label{c13}
\end{equation}%
where the function $\Phi _{k}\left( z\right) $ satisfies the
following
equation%
\begin{equation}
-\frac{d}{dz}\left( z^{2}\frac{d\Phi _{k}\left( z\right)
}{dz}\right) +ikL_{c}\Phi _{k}\left( z\right) +z\Phi _{k}\left(
z\right) =\Psi \left( z\right)   \label{c14}
\end{equation}%
with the length $L_{c}$ equal to%
\begin{equation}
L_{c}=2\pi \nu SD_{0}  \label{c15}
\end{equation}%
The length $L_{c}$ is actually the localization length, which will
be seen from the final result. Equations (\ref{c12}-\ref{c14}) can
also be obtained for the orthogonal and symplectic ensembles but
with different localization lengths $L_{c}$. The result can be
written as
\begin{equation}
L_{c}^{symplectic}=2L_{c}^{unitary}=4L_{c}^{orthogonal}
\label{c16}
\end{equation}
The residue of the function $K_{\omega }$ is proportional to the function $%
p_{\infty }\left( r,r^{\prime },\varepsilon \right) $ introduced
by Anderson
\cite{anderson},%
\begin{equation}
p_{\infty }\left( r,r^{\prime },\varepsilon \right)
=\sum_{k}\left\vert \phi _{k}\left( r\right) \right\vert
^{2}\left\vert \phi _{k}\left( r^{\prime }\right) \right\vert
^{2}\delta \left( \varepsilon -\varepsilon _{k}\right) ,
\label{c17}
\end{equation}%
where $\phi _{k}\left( r\right) $ are exact eigenfunctions.

Eqs. (\ref{c12}-\ref{c14}) exactly coincide with the low frequency
limit of equations derived by Berezinsky \cite{berezinsky}
provided the length $L_{c}$ is replaced by the mean free path $l,$
which shows that the low frequency limit of the one dimensional
systems is universal.

The exact solution of Eqs. (\ref{c12}-\ref{c14}) leads to the
following
expression%
\begin{equation}
p_{\infty }\left( x\right) =\frac{\pi ^{2}\nu
}{16L_{c}}\int_{0}^{\infty
}\left( \frac{1+y^{2}}{1+\cosh \pi y}\right) ^{2}\exp \left( -\frac{1+y^{2}}{%
4L_{c}}\left\vert x\right\vert \right) y\sinh \pi ydy  \label{c18}
\end{equation}
In the limit $x\gg L_{c}$, Eq. (\ref{c18}) reduces to a simpler form%
\begin{equation}
p_{\infty }\left( x\right) \thickapprox \frac{\nu }{4\sqrt{\pi
}L_{c}}\left(
\frac{\pi ^{2}}{8}\right) ^{2}\left( \frac{4L_{c}}{\left\vert x\right\vert }%
\right) ^{3/2}\exp \left( -\frac{\left\vert x\right\vert
}{4L_{c}}\right) \label{c19}
\end{equation}
The exponential form of $p_{\infty }\left( x\right) $ proves the
localization of the wave functions and shows that the length
$L_{c}$ is the
localization length. Note, however, the presence of the pre-exponential $%
\left\vert x\right\vert ^{-3/2}$. Due to the factor the integral over $x$ of $%
p_{\infty }\left( x\right) $ remains finite even in the limit $%
L_{c}\rightarrow \infty $. Actually, one obtains%
\begin{equation}
\int_{-\infty }^{\infty }p_{\infty }\left( x\right) dx=\nu ,
\label{c20}
\end{equation}%
which proves the localization of all states.

At small $k\ll L_{c}^{-1}$, the function $A\left( k\right) $ in Eq. (\ref%
{c13}) takes the form%
\[
A\left( k\right) =1-4\zeta \left( 3\right) k^{2}L_{c}^{2}
\]%
and the static dielectric permeability $\epsilon $ equals
\begin{equation}
\epsilon =-4\pi e^{2}\nu \left. \frac{d^{2}A\left( k\right) }{dk^{2}}%
\right\vert _{k=0}=32\zeta \left( 3\right) e^{2}\nu L_{c}^{2}
\label{c21}
\end{equation}%
where $\zeta \left( x\right) $ is the Riemann $\zeta $-function.

All these calculations have been performed for a finite frequency
$\omega $ and the infinite length of the sample. One can also
consider the case of the zero frequency and a finite length $L$. A
full analysis of this limit has been presented by
Zirnbauer\cite{zirnbauer} who calculated the average conductivity
as a function of $L$.

There is another Fokker-Planck approach to study transport of
disordered wires developed by Dorokhov, Mello, Pereyra, and Kumar
\cite{dorokhov,mpk} (DMPK method). It can be applied also to thin
wires with a small number of channels. At the same time, this
method cannot be used for finite frequencies. In the case of thick
wires with a large number of the channels and zero frequencies,
the equivalence of the supersymmetry to the DMPK method has been
demonstrated by Brouwer and Frahm \cite{bf}.

Many interesting problems of banded random matrices \cite{fm} and
quantum
chaos (like kicked rotor\cite{az1}) can be mapped onto the $1D$ supermatrix $%
\sigma $-model. However, a detailed review of these interesting
directions of research is beyond the scope of this paper.

\section{Anderson localization in $2$ and $2+\protect\epsilon $
dimensions.}

Study of localization in $2$ and $2+\epsilon $ using the replica $\sigma $%
-model was started by Wegner \cite{wegner} using a renormalization
group (RG) technique. He was able to write the RG equations for
the orthogonal and
unitary ensembles that could be used in $2$ dimensions and extended into $%
2+\epsilon $ dimensions for $\epsilon \ll 1$. The latter was done
with a hope that putting $\epsilon =1$ at the end of the
calculations one could extract at least qualitatively an
information about the Anderson metal-insulator transition in $3$
dimensions. Based on this calculation a conclusion about the
localization at any weak disorder in $2D$ was made. As concerns
$2+\epsilon $, an unstable fixed point was found, which following
the standard arguments by Polyakov \cite{polyakov} signaled the
existence of the metal-insulator transition.

The symplectic case was considered within the compact replica
$\sigma $-model in Ref. \cite{elk} using the same method of RG and
it was shown that the resistivity had to vanish in the limit of
$\omega \rightarrow 0$. The
difference between the replica $\sigma $-models  used in Refs. \cite%
{wegner,sw} (noncompact) and Ref.\cite{elk} (compact) is not
essential when applying the RG scheme.

Exactly the same results are obtained with the supermatrix $\sigma
$-model using the RG technique \cite{efetovrg,efetovreview,book}
and let me sketch the derivation here. As usual in the RG method,
one introduces a running
cutoff parameter and coupling constants depending on this cutoff. The $%
\sigma $-model for such couplings can be written as
\begin{equation}
F=\frac{1}{t}\int STr\left[ \left( \nabla Q\right) ^{2}+2i\tilde{\omega}%
\Lambda Q\right] dr  \label{d1}
\end{equation}%
where $\tilde{\omega}=\omega /D_{0}$. The bare value of $t$ equals $%
t=8\left( \pi \nu D_{0}\right) ^{-1}$ (c.f. Eq. (\ref{a24})).

The $\sigma $-model looks similar to classical spin $\sigma
$-models considered in Ref. \cite{polyakov} and one can follow the
RG procedure suggested in that work. Using the constraint
(\ref{a23}) one can write the
supermatrix $Q$ in the form%
\begin{equation}
Q=V\Lambda \bar{V},  \label{d2}
\end{equation}%
where $V\bar{V}=1$ so that $V$ is a pseudo-unitary supermatrix.

In order to integrate over a momentum shell one can represent the
supermatrix $V$ in the form%
\begin{equation}
V\left( r\right) =\tilde{V}\left( r\right) V_{0}\left( r\right) ,
\label{d3}
\end{equation}%
where $V_{0}$ is a supermatrix fast varying in space and
$\tilde{V}$ is slowly varying one. These supermatrices have the
same symmetry as the supermatrix $V$.

Substituting Eq. (\ref{d3}) into Eq. (\ref{d1}) one can write the
free
energy functional $F\left[ Q\right] $ in the form%
\begin{equation}
F=\frac{1}{t}\int STr\left[ \left( \nabla Q_{0}\right)
^{2}+2\left[
Q_{0},\nabla Q_{0}\right] \Phi +\left[ Q_{0},\Phi \right] ^{2}+2i\tilde{%
\omega}\overline{\tilde{V}}\Lambda \tilde{V}Q_{0}\right] dr
\label{d4}
\end{equation}

\begin{equation*}
Q_{0}=V_{0}\Lambda \bar{V}_{0},\qquad \Phi
=\overline{\tilde{V}}\nabla \tilde{V}=-\bar{\Phi}
\end{equation*}%
The next step of the RG procedure is to integrate over the fast
varying matrices $Q_{0}$ and reduce to a functional containing
only slowly varying variables $V$. After this integration the free
energy $F$ in Eq. (\ref{d4})
should be replaced by energy $\tilde{F}$ describing the slow fluctuations%
\begin{equation}
\tilde{F}=-\ln \int \exp \left( -F\right) DQ_{0}  \label{d5}
\end{equation}%
The integration over the supermatrix $Q_{0}$ can be done using a
parametrization (\ref{a25a}) or a more convenient parametrization
\begin{equation}
Q_{0}=\Lambda \left( 1+P\right) \left( 1-P\right) ^{-1},\text{\qquad }%
P\Lambda +\Lambda P=0.  \label{d6}
\end{equation}%
Integration over the fast variation means that one integrates over
Fourier transformed $P_{k}$ with $\lambda k_{0}<k<k_{0}$, where
$k_{0}$ is the upper cutoff and $\lambda <1$. As a result of the
integration one comes to the
same form of the functional $F$ as in Eq. (\ref{d1}). The constant $\tilde{%
\omega}$ does not change under the renormalization but the new
coupling
constant $\tilde{t}$ can be written as%
\begin{equation}
\tilde{t}^{-1}=t^{-1}\left( 1+\frac{\alpha t}{8}\int_{\lambda k_{0}}^{k_{0}}%
\frac{d^{d}k}{k^{2}\left( 2\pi \right) ^{d}}\right)   \label{d7}
\end{equation}%
The correction to the coupling constant $t$, Eq. (\ref{d7}), is
written in the first order in $t$. The parameter $\alpha $ depends
on the ensemble and
equals%
\begin{equation}
\alpha =\left\{
\begin{array}{cc}
-1, & orthogonal \\
0, & unitary \\
1 & symplectic%
\end{array}%
\right.   \label{d8}
\end{equation}%
Stretching the coordinates in the standard way and changing the
notation for the coupling constant $t\rightarrow 2^{d+1}\pi
d\Gamma \left( d/2\right) t$, where $\Gamma $ is the Euler $\Gamma
$-function one obtains the RG equation for $t$
\begin{equation}
\beta \left( t\right) =\frac{dt}{d\ln \lambda }=\left( d-2\right)
t+\alpha t^{2}  \label{d9}
\end{equation}
where $\beta \left( t\right) $ means the Gell-Mann-Low function.

In $2D$, the solution of this equation for the coupling constant
$t$ (proportional to resistivity) takes the form
\begin{equation}
t\left( \omega \right) =\frac{t_{0}}{1+\alpha t_{0}\ln \left(
1/\omega \tau \right) }  \label{d10}
\end{equation}%
For sufficiently high frequencies $\omega $ the resistivity and
the
diffusion coefficient $D\left( \omega \right) $ proportional to $%
t^{-1}\left( \omega \right) $ coincide with their bare values.

Decreasing the frequency $\omega $ results in growing the
resistivity for the orthogonal ensemble until the coupling
constant $t\left( \omega \right) $ becomes of the order $1$. Then,
the RG scheme is no longer valid because the
expansion in $t$ in the R.H.S. of Eq. (\ref{d9}) is applicable only for $%
t\ll 1$. However, it is generally believed that $t$ diverges in the limit $%
\omega \rightarrow 0$ and this should mean the localization of all
states with an exponentially large localization length
\begin{equation}
L_{c}\propto \exp \left( 1/t_{0}\right)   \label{d11}
\end{equation}

In the symplectic ensemble the resistivity $t\left( \omega \right)
$ decreases with decreasing the frequency $\omega $. This
interesting result was obtained in the first order in $t_{0}$ by
Hikami, Larkin and Nagaoka \cite{hln}. However, Eq. (\ref{d10})
means more\cite{elk}. If the bare $t_{0} $ is small, $t_{0}\ll 1$,
the effective resistivity $t\left( \omega \right) $ decays down to
zero in the limit $\omega \rightarrow 0$. In this case the
constant $t\left( \omega \right) $ is small for any frequency and
the one loop approximation used in the derivation of Eq.
(\ref{d9}) is valid for all frequencies. So, the solution for the
symplectic ensembles, when used for the low frequencies, is the
most reliable one obtained with the RG method.

As concerns the unitary ensemble, the first order contribution
vanishes and one should calculate corrections of the second order.
As a result, one comes to the following dependence of $t\left(
\omega \right) $ on the
frequency%
\begin{equation}
t\left( \omega \right) =\frac{t_{0}}{\left( 1-t_{0}^{2}\ln \left(
1/\omega \tau \right) \right) ^{1/2}}  \label{d12}
\end{equation}%
One can see from Eq. (\ref{d12}) that the resistivity $t\left(
\omega
\right) $ grows, as in the orthogonal ensemble, until it becomes of order $1$%
. Again, this behavior is interpreted as localization for any
disorder. The conclusions about the localization in $2D$ for the
orthogonal and unitary ensembles were made first in Ref.
\cite{wegner} and this agreed with the results based on using the
scaling hypothesis \cite{aalr}.

Wegner developed also theory of the Anderson metal-insulator
transition
in the dimensionality $2+\epsilon $ for $\epsilon \ll 1$ \cite{wegner0}%
. One can see that the RG equation (\ref{d9}) has a fixed point $%
t_{c}=\epsilon ,$ at which the Gell-Mann-Low function vanishes. At
this point the total resistance of the sample does not depend on
the sample size and this point should correspond to the Anderson
metal-insulator transition.

Linearizing function $\beta \left( t\right) $ near the fixed point
$t_{c}$ one can solve Eq. (\ref{d9}). As a result one can find a
characteristic
(correlation) length $\xi $ near the fixed point%
\begin{equation}
\xi \sim \xi _{0}\left( \frac{t_{c}-t_{0}}{t_{c}}\right)
^{-1/y},\quad y=-\beta ^{\prime }\left( t_{c}\right)   \label{d13}
\end{equation}%
where $\xi _{0}$ is the size of a sample having the entire resistance $t_{0}$%
. Assuming that the length $\xi $ is the only characteristic
length in the system and that the conductivity $\sigma $ is
proportional to $t_{c}^{-1}\xi
^{2-d}$, one can write the equation for the conductivity in the following form%
\begin{equation}
\sigma =A\frac{e^{2}}{\xi _{0}^{d-2}t_{c}}\left( \frac{t_{c}-t}{t_{c}}%
\right) ^{s},\quad s=\frac{d-2}{y}  \label{d14}
\end{equation}

The explicit values of the critical resistance $t_{c}$ and the
exponent $s$ for the orthogonal and unitary ensembles equals
\begin{equation}
\tilde{t}_{c}=\left\{
\begin{array}{cc}
d-2, & orthogonal \\
\left( 2\left( d-2\right) \right) ^{1/2}, & unitary%
\end{array}%
\right.   \label{d15}
\end{equation}%
and%
\begin{equation}
s=\left\{
\begin{array}{c}
1 \\
1/2%
\end{array}%
\right.   \label{d16}
\end{equation}

Eqs. (\ref{d13}-\ref{d16}) demonstrate that the metal-insulator
transition exists in any dimensionality $d>2$ and the conductivity
near the transition
obeys a power law. Of course, this consideration is restricted by small $%
\epsilon =d-2$ and one can use the result in $3D$ only
qualitatively.

The scaling approach developed for small $\epsilon $ is similar to
the one developed for conventional phase transitions in, e.g.,
spin models where one can also write $\sigma $-models. This method
is not sensitive to whether the symmetry of the supermatrices $Q$
is compact or noncompact. Using this approach one comes to the
conclusion that the Anderson metal-insulator transition is very
similar to standard second order phase transitions.

In the next section the same problem will be considered on the
Bethe lattice or in a high dimensionality. Surprisingly, the
result will be very different and the peculiarity of the solution
originates from the noncompactness of the group of the symmetry of
the supermatrices $Q$.

\section{Anderson metal-insulator transition on the Bethe lattice or in a
high dimensionality.}

It is generally difficult to find the critical point for a
transition between different states and describe the critical
behavior in its vicinity. The Anderson metal-insulator transition
is definitely not an exception in this respect. Usually,
identifying a proper order parameter one can get an idea about a
transition using a mean field approximation. As concerns the
Anderson transition, this is not possible. Although the $\sigma
$-model, Eq. (\ref{a24}), looks very similar to spin models in a
magnetic field, one
cannot take an average of $Q$ with the free energy $F\left[ Q\right] ,$ Eq. (%
\ref{a24}), as the order parameter because it determines the
average density of states and is not related to the Anderson
transition.

At the same time, the mean field approximation works very well in
high dimensionality or on special structures like the Bethe
lattice.

The Anderson model of the Bethe lattice was studied for the first
time by Abou-Chacra, Anderson and Thouless \cite{aat}, who proved
the existence of the metal-insulator transition and found the
position of the mobility edge. With the development of the
supersymmetry technique it became possible to describe the
critical behavior both in the metallic and insulating regime.
Considering a granular model one could obtain results for the
orthogonal, unitary and symplectic ensembles. Later the Anderson
model has also been described.

It turned out that in all the cases the critical behavior was the
same, which contrasts the results obtained within the $2+\epsilon
$ expansion. This could not be a big surprise because for most
phase transitions the high dimensional results are more
\textquotedblleft universal" than those obtained in lower
dimensions. However, the results for the metallic and insulating
regimes did not obey the conventional scaling and this was
completely unexpected.

The first attempt to solve the granular version of the supermatrix $\sigma $%
-model on the Bethe lattice has been undertaken in Ref.
\cite{efetov84}. In this work correct integral equation have been
written for description of critical behavior near the
metal-insulator transition and the position of the mobility edge
has been found. However, attempts to find a solution of this
equation related to scaling properties of the $2+\epsilon $ limit
were not successful, which lead to wrong conclusions.

Studying numerically the integral equation derived in
\cite{efetov84} Zirnbauer\cite{zirnbauerbethe} found a very
unusual behavior near the critical point and presented formal
reasons explaining this behavior. Finally, the density-density
correlation function has been calculated for
the unitary \cite{efetov87} and orthogonal and symplectic ensembles \cite%
{efetov87}. This determined the diffusion coefficient in the
metallic region and localization length and dielectric
permeability in the insulating one.

The form of the density-density correlation function on the Bethe
lattice differs from the one on conventional lattices. Therefore
the problem of the Anderson localization has been considered on
such lattices in an effective medium approximation
\cite{efetov88}. The latter becomes exact on the real lattices in
a high dimensionality $d\gg 1$ and the basic equations and results
are similar. The derivation of the equations and the final results
are shortly displayed below. A detailed discussion can be found in \cite%
{book}

The scheme of the derivation of the equations is similar to the
one presented in Sec. \ref{wires} for one dimensional structures
consisting of the grains. We start with Eq. (\ref{c2}) written on
a $d$-dimensional lattice with $d\gg 1$ or on the Bethe lattice.
Denoting by $\Psi \left( Q\right) $ the partition function of a
branch of the tree structure with a fixed value $Q$ at the base
and comparing it with the partition function on the neighboring
site one comes to a non-linear integral equation
\begin{equation}
\Psi \left( Q\right) =\int N\left( Q,Q^{\prime }\right)
Z_{0}\left( Q^{\prime }\right) \Psi ^{m}\left( Q^{\prime }\right)
dQ^{\prime } \label{e1}
\end{equation}
where $m=2d-1$ for a $d$-dimensional lattice and is the branching
number on
the Bethe lattice. The functions $N\left( Q,Q^{\prime }\right) $ and $%
Z_{0}\left( Q\right) $ have been introduced in Eq. (\ref{c6}).

The case $m=1$ corresponds to the one-dimensional chains of the
grains and Eq. (\ref{e1}) coincides with Eq. (\ref{c5}) in this
limit. In this
particular case equation (\ref{e1}) is linear and, as we have seen in Sec. %
\ref{wires}, all states are localized for any disorder. However,
at $m>1$ the integral equation (\ref{e1}) is non-linear and has a
bifurcation at a critical $\alpha _{c}$ corresponding to the
Anderson metal-insulator transition.

The density-density correlation function $K_{\omega }$, Eq.
(\ref{a25}), can
be written in the form%
\begin{equation}
K_{\omega }\left( r_{1},r_{2}\right) =-2\pi ^{2}\nu
\tilde{\nu}\int Q_{33}^{12}P_{33}\left( r,Q\right) Z\left(
Q\right) \Psi \left( Q\right) dQ \label{e2}
\end{equation}%
where the function $P\left( r,Q\right) $ satisfies for the high
dimensional
lattices the following equation%
\begin{eqnarray}
&&P\left( r,Q\right) -\sum_{r^{\prime }}W\left( r-r^{\prime
}\right) \int N\left( Q,Q^{\prime }\right) P\left( r^{\prime
},Q^{\prime }\right) Z\left(
Q^{\prime }\right) dQ^{\prime }  \label{e3} \\
&+&m\int N_{2}\left( Q,Q^{\prime }\right) P\left( r,Q^{\prime
}\right) Z\left( Q^{\prime }\right) dQ^{\prime } =\delta \left(
r\right) Q^{21}\Psi \left( Q\right) .  \notag
\end{eqnarray}

In Eq. (\ref{c3}) the function $N_{2}\left( Q,Q^{\prime }\right) $
is equal to
\begin{equation*}
N_{2}\left( Q,Q^{\prime }\right) =\int N\left( Q,Q^{\prime \prime
}\right) N\left( Q^{\prime \prime },Q\right) Z\left( Q^{\prime
\prime }\right) dQ^{\prime \prime }
\end{equation*}%
and%
\begin{equation*}
W\left( r-r^{\prime }\right) =\left\{
\begin{array}{cc}
1, & \left\vert r-r^{\prime }\right\vert =1 \\
0, & \left\vert r-r^{\prime }\right\vert \neq 1%
\end{array}%
\right.
\end{equation*}%
The third term in the L.H.S. of Eq. (\ref{e3}) takes into account
the fact
that two segments of a broken line cannot coincide. Eqs. (\ref{e2}, \ref{e3}%
) are very similar to Eqs. (\ref{c4}, \ref{c7a}) written for the
$1D$ case. This is natural because in both the cases loops are
absent. Their solution for a function $\Psi \left( Q\right) $
found from Eq. (\ref{e1}) can be obtained making a spectral
expansion of $P\left( r,Q\right) $ in
eigenfunctions of the integral operators entering the L.H.S. of Eq. (\ref{e3}%
).

In principle, this procedure is straightforward. However, solving
the integral equation (\ref{e1}) is not simple because it contains
a large number of the elements of the supermatrix $Q$.

Fortunately, Eqs. (\ref{e1}-\ref{e3}) drastically simplify in the
metallic regime near the metal-insulator transition and
everywhere in the insulating regime provided one considers the low
frequency limit $\omega \rightarrow 0$. The formal reason for this
simplification is that the main contribution into the correlation
functions comes in these cases from the region of very large
values of the variables $\lambda _{1}\gtrsim \Delta /\omega \gg
1$. The same simplification has helped one to solve the problem of
the localization in wires in Sec. \ref{wires}.

Nevertheless, the full analysis is quite involved even for small
$\omega $. Details can be found again in Ref. \cite{book} and here
I display only the final results.

\begin{figure}
\centerline{\psfig{file=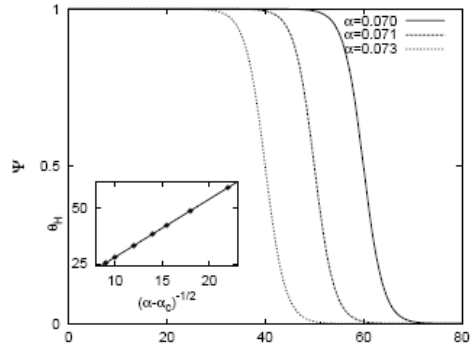,width=7.2cm}} \caption{Numerical
solution $\Psi(\lambda_1)$ for $m=2$ and different hopping
amplitudes in the critical metallic regime. The inset shows
$\theta_H$ defined by $\Psi(cosh\theta_H)=0.5$ as a function of
$(\alpha-\alpha_c)^{-1/2}$} \label{fig2}
\end{figure}

In the insulating regime, $\alpha <\alpha _{c}$, only $\Psi =1$ is
the solution of Eq. (\ref{e1}) in the limit $\omega =0$. This
solution of the simplified equation persists for all $\alpha $ but
another solution appears in the region $\alpha >\alpha _{c}$. The
latter solution considered as a function of $\theta _{1}=\ln
\left( 2\lambda _{1}\right) $ has a form of a
kink moving to infinity as $\alpha \rightarrow \alpha _{c}$. The position $%
\theta _{1c}$ of the kink depends on the distance from the critical point $%
\alpha _{c}$ as%
\begin{equation}
\theta _{1c}=s\left( \alpha -\alpha _{c}\right) ^{-1/2}
\label{e4}
\end{equation}%
where $s$ is a number of order $1$.  The dependence of $\Psi
\left( \lambda _{1}\right) $ is represented in Fig. \ref{fig2}

Only this solution should be used for $\alpha >\alpha _{c}$ and
this leads to a very non-trivial critical behavior of the
diffusion coefficient.

The position of the critical point $\alpha _{c}$ and the critical
behavior have been calculated for all $3$ ensembles. For large
$m$, the value $\alpha _{c}$ for the orthogonal and unitary
ensembles is determined by the following
equations%
\begin{equation}
\begin{array}{cc}
\frac{2^{3/2}}{\pi }\left( \frac{\alpha _{c}}{2\pi }\right) ^{1/2}m\ln \frac{%
\gamma }{\alpha _{c}}=1, & orthogonal \\
\left( \frac{\alpha _{c}}{2\pi }\right) ^{1/2}m\ln \frac{2}{\alpha
_{c}}=1,
& unitary%
\end{array}
\label{e5}
\end{equation}%
One can see from Eq. (\ref{e5}) that the metallic region is
broader for systems with the broken time reversal invariance. In
other words, applying a
magnetic field shifts the metal-insulating transition to larger values of $%
\alpha _{c}.$ This result correlates with the one, Eq.
(\ref{d15}), obtained in $2+\epsilon $ dimensions.

Although the position of the Anderson transition depends on the
ensemble considered, the form of the correlation functions is the
same.

In the insulating regime, the function $p_{\infty }\left( r\right) $, Eq. (%
\ref{c17}), takes for $r\gg L_{c}$ the following form%
\begin{equation}
p_{\infty }\left( r\right) =const\,r^{-\left( d+2\right)
/2}L_{c}^{-d/2}\exp \left( -\frac{r}{4L_{c}}\right)   \label{e6}
\end{equation}%
where $L_{c}$ is the localization length.

Near the transition the localization length $L_{c}$ grows in a power law%
\begin{equation}
L_{c}=\frac{const}{\left( \alpha _{c}-\alpha \right) ^{1/2}}
\label{e7}
\end{equation}%
In this regime there is another interesting region of $1\ll r\ll
L_{c}$
where the function $p_{\infty }\left( r\right) $ decays in a power law%
\begin{equation}
p_{\infty }\left( r\right) =const\,r^{-d-1}  \label{e8}
\end{equation}%
Remarkably, Eq. (\ref{e6}) obtained for $d\gg 1$ properly
describes also the one-dimensional wires (c.f. Eq. (\ref{c19})).

The integral of $p_{\infty }\left( r\right) $ over the volume is
convergent for all $\alpha \leq \alpha _{c}$ and remains finite in
the limit $\alpha \rightarrow \alpha _{c}$ indicating that the
wave functions at the transition point decay rather fast. At the
same time, all moments of this quantity diverge in this limit. The
second moment determines the
electric susceptibility $\kappa $,%
\begin{equation}
\kappa \delta _{\alpha \beta }=e^{2}\int r_{\alpha }r_{\beta
}p_{\infty }\left( r\right) d^{d}r  \label{e9}
\end{equation}
Near the transition calculation of the integral in Eq. (\ref{e9})
leads to
the result%
\begin{equation}
\kappa =4\pi ^{2}\nu cL_{c}  \label{e10}
\end{equation}%
where $c$ is a coefficient.

This equation shows that the susceptibility in the critical region
is proportional to the localization length $L_{c}$ and not to
$L_{c}^{2}$ as it would follow from the one-parameter scaling
\cite{aalr} and results obtained
in $2+\epsilon $ dimensions. The unusual dependence of the susceptibility $%
\kappa $ on $L_{c}$ in Eq. (\ref{e10}) arises formally from the
anomalous exponent $\left( d+2\right) /2$ in the power law
behavior of the pre-exponent in Eq. (\ref{e6}).

As concerns the metallic regime one comes for the real lattices to
the diffusion propagator, Eq. (\ref{a25b}), at all $\alpha >\alpha
_{c}$. However, except the limit $\alpha \gg 1$, the diffusion
coefficient $D$ obtained now is different form the classical
diffusion coefficient $D_{0}$. Its behavior in the critical region
$\alpha -\alpha _{c}\ll \alpha _{c}$ is especially interesting.
This is not a power law behavior as one could expect from the one
parameter scaling. Instead, the diffusion coefficient decays
near the transition exponentially%
\begin{equation}
D=const\frac{\exp \left[ -s\left( \alpha -\alpha _{c}\right)
^{-1/2}\right] }{\left( \alpha -\alpha _{c}\right) ^{3/2}}
\label{e11}
\end{equation}%
This is a very unusual behavior. Formally, it follows from the
non-compact symmetry of the supermatrices $Q.$ For any compact
symmetry one would obtain in the same approximation a power law
dependence of the diffusion coefficient on $\alpha -\alpha _{c}$.
The exponential decay of the diffusion
coefficient $D$, Eq. (\ref{e11}), follows from the shape of the function $%
\Psi $, Fig. \ref{fig2}. The position of the kink $\lambda
_{H}=\cosh \theta _{H}$, Eq. (\ref{e7}), goes to infinity as
$\lambda _{H}\propto \exp \left[ s\left( \alpha -\alpha
_{c}\right) ^{1/2}\right] $ and this results in the form
(\ref{e11}) of the diffusion coefficient.

The same results, Eqs. (\ref{e6}-\ref{e11}), have been obtained
later \cite{mirfyod} for the Anderson model on the Bethe lattice
and this completed the study of this model started in Ref.
\cite{aat}. The agreement of the results obtained for the Anderson
model and granulated $\sigma $-model is, of course, not accidental
because, the critical behavior is formed by long time correlations
and the result should not be sensitive to short distance
structures. As it has been discussed previously, the low frequency
behavior of wires and strictly one dimensional chains is also
described by identical equations.

The exponential decay of the diffusion coefficient was interpreted
in Ref. \cite{book} in terms of tunnelling between quasi-localized
states. This may happen provided the wave function is concentrated
in centers with a large distance
\begin{equation}
\zeta \propto \left( \alpha -\alpha _{c}\right) ^{-1/2}
\label{e12}
\end{equation}
between them. The decay of the amplitudes of the wave functions in
the
single center is fast as it can be seen from the fast decay in Eq. (\ref{e8}%
). Then, the tunnelling leads to an overlap between the wave
functions of the different centers and to formation of a
conduction band with an
effective bandwidth $\Gamma ,$%
\begin{equation}
\Gamma \propto \Delta \exp \left( -a\zeta \right)  \label{e13}
\end{equation}
where $a$ is a coefficient.

The exponential decay of the diffusion coefficient $D$, Eq.
(\ref{e11}), can follow quite naturally from such a picture. Of
course, the picture implies the existence of weakly overlapping
centers of the localization near this transition. A well
established multifractality of wave functions at the transition
(for a recent review, see, e.g. \cite{evers}) may point out on
this strong inhomogeneity near the transition.

Another indication in favor of the presented picture comes from
the fact that the solution $\Psi $ of Eq. (\ref{e1}) looses the
sensitivity to the existence of the transition at frequencies
$\omega \gtrsim $ $\Gamma $. This can be seen from a more detailed
analysis of Eq. (\ref{e1}). The interpretation in terms of
formation of a very narrow conduction band near the transition
with the bandwidth $\Gamma $ is consistent with this property of
the solution $\Psi $.

The fixed point found in $2+\epsilon $ for small $\epsilon $
corresponds to a weak disorder and the strong inhomogeneities are
not seen in this approach. One cannot speak of a narrow conduction
band near the transition in $2+\epsilon $ dimensions within this
picture.

In principle, centers of (quasi) localization exist in $2D$ and
can be described in the framework of $\sigma $-model (for a
review, see Refs. \cite{book,mirlin}). However, the idea about
these centers
of the (quasi) localizations is not incorporated in the conventional $%
2+\epsilon $ scheme. So, the standard continuation of the results
obtained for small $ \epsilon $ to $\epsilon =1$ may result in
loosing an important information.

The non-trivial form of the function $\Psi $ (see e.g. Fig.
\ref{fig2}) has lead the present author to the idea
\cite{efetov88} that this function might play the role of an order
parameter for the Anderson transition. It was guessed that a
Laplace transform of this function could be related to a
conductance distribution. This idea has been further developed in Ref. \cite%
{efetov90}, where a functional in an extended space was
constructed such that its minimum was reached at the function
$\Psi \left( Q\right) $. This resembles the Landau theory of phase
transitions but the role of the order parameter is played a by a
function.

The concept of the function order parameter was also discussed in
later works on the Bethe lattice \cite{mirfyod}.

\section{Discussion.}

In this paper the basics of the supersymmetry method has been
presented. It is explained how the non-linear supermatrix $\sigma
$-model is derived and it is shown how one can calculate within
this model. It is demonstrated how one comes to the Wigner-Dyson
statistics in a limited volume and how one obtains Anderson
localization in disordered wires. Renormalization group scheme is
explained in $2$ and $2+\epsilon $ dimensions for small $\epsilon
,$ renormalization group equations are written and solved. It is
shown how one solves the problem of the Anderson metal-insulator
transition on the Bethe lattice and high dimensionality.

From the technical point of view all this was a demonstration how
one can calculate in the dimensions $d=0$, $d=1$, $d=2,$ and $d\gg
1$. Due to the lack of the space the present paper is not a
complete review of the application of the supersymmetry technique
and many interesting works are not mentioned. However, the
calculational schemes presented here have been used in most of the
subsequent works. So, having read this paper one can get an idea
on how one can work in all situations where the supersymmetry
method is useful.

This is a chapter in the book devoted to 50 years of the Anderson
localization and I tried to describe shortly how one of the
directions of the field was developing in
1980s after the second most important work on the Anderson localization \cite%
{aalr} has been published. Many of the authors of the present
volume entered this field motivated by this publication. I hope
that the development of the supersymmetry method has been useful
in solving several interesting problems of the Anderson
localization.

Although the supersymmetry method proved to be an adequate method
for studying disordered systems (at least, without
electron-electron interaction), several very important problems
have not been solved so far.
 In spite of the common believe that
all states are localized in disordered films (orthogonal and
unitary ensembles), the solution for the two-dimensional $\sigma
$-model has not been found in the limit of low frequencies. The
problem of the integer quantum Hall effect has not been solved
either, although the idea about instantons \cite{pruisken} was
very useful for the understanding of this phenomenon. The problem
of describing the critical behavior near the transition between
the Hall plateaus still awaits its resolution.

One more interesting problem is to
understand the critical behavior near the Anderson transition.

Of course, a lot of information comes from numerical simulations
but solving the $2D$ problem analytically would be really a great
achievement. As concerns the Anderson transition in $3D,$ the hope
to solve it exactly is not realistic because even a simpler Ising
model has not been solved in spite of numerous attempts. However,
in the conventional theory of phase transition one can start with
a mean field theory justifiable in high dimensions, determine the
upper critical dimension and then make an expansion near this
dimensionality.

Unfortunately, until now a similar procedure has not been found
for the Anderson transition, although the supermatrix $\sigma
$-model resembles spin models for which this procedure is
standard. This concerns also the $2D$ case, where conventional
spin $\sigma $-models are solvable. However, the well developed
methods like the Bethe Ansatz or methods of the conformal field
theory do not work here.

The formal reason of the failure of these approaches for studying
the supermatrix $\sigma $-model is that the group of the symmetry
of the supermatrices $Q$ is not compact. These supermatrices
consist of a block varying on a sphere and another one with
elements on the hyperboloid. The latter part of $Q$ is formally
responsible for the localization but its presence leads to
difficulties when applying the well developed methods. It is clear
that the importance of the noncompact symmetry is not fully
appreciated.

I can only express my hope that these problems will be
resolved in the next 50 years and the book devoted to 100 years of
the Anderson localization will contain the complete theory of this
phenomenon.

\section*{Acknowledgements}

The work was supported by Transregio 12 of DFG ``Symmetries and
Universality in Mesoscopic Systems"

\end{document}